\documentstyle[12pt]{article}
\newcommand{\bq}{\begin{equation}}
\newcommand{\eq}{\end{equation}}
\newcommand{\ba}{\begin{eqnarray}}
\newcommand{\ea}{\end{eqnarray}}

\newcommand{\nl }{ \nonumber  }
\newcommand{\vf}{\varphi}
\newcommand{\ul}{\underline}
\newcommand{\p}{\partial}
\newcommand{\pu}{\p_\tau}
\newcommand{\pa}{\p_a}
\newcommand{\h}{\hspace{1cm}}
\begin{document}
\pagestyle{empty,textwidth,textheight}
\begin{titlepage}
\begin{flushright}
hep-th$/9711136$
\\
JINR-E2-98-84
\end{flushright}
\vspace*{2cm}
{\bf\begin{center}
 TENSIONLESS BRANES AND THE NULL STRING CRITICAL DIMENSION
\footnote{Work supported in part by the National Science Foundation
of Bulgaria under contract $\phi-620/1996$}
\vspace*{2cm}
\\
P. Bozhilov
\footnote {permanent address:
\it Dept.of Theoretical Physics,"Konstantin Preslavsky" Univ. of Shoumen
\\
9700 Shoumen, Bulgaria  E-mail: bojilov@uni-shoumen.bg}
\\
\it
Bogoliubov Laboratory of Theoretical Physics, JINR, Dubna, Russia
\\
\it
E-mail: bojilov@thsun1.jinr.dubna.su
\vspace*{5cm}
\end{center}}

      BRST quantization is carried out for a model of $p$-branes with
second class constraints. After extension of the phase space the
constraint algebra coincides with the one of null string when $p=1$.
It is shown that in this case one can or can not obtain critical
dimension for the null string, depending on the choice of the operator
ordering and corresponding vacuum states. When $p>1$, operator orderings
leading to critical dimension in the $p=1$ case are not allowed. Admissable
orderings give no restrictions on the dimension of the embedding space-time.
Finally, a generalization to supersymmetric null branes is proposed.
\end{titlepage}
\normalsize
\vspace*{2cm}
\section{\bf Introduction}

Recently a lot of papers are devoted to the tensionless (null) strings and
their application in different areas and different dimensions \cite{A}.
In connection with the above activity it is worth to consider again
the question about the critical dimension of the null strings and more
generally - null $p$-branes. There are two answers to this question in the
literature. Most of the authors insist on the nonexistence of critical
dimension for such objects \cite{B}, but some other receive opposite
results \cite{C}.
In our opinion the reason is in the different approaches to the choice of
the operator ordering. If one looks at the classical null string as a
collection of particles moving under certain conditions and wants to keep
this picture in the quantum case also, there is no reason to expect a
critical dimension emerging. Therefore one adopts such operator ordering
which supports this point of view. If one does not bother about previous
particle interpretation but simply compares the appearance of the anomaly in
the constraint algebras of the null and usual string upon quantization
(with one and the same operator ordering), one
sees that the non-trivial central terms arise independently of the string
tension $T$. Then the existence of critical dimension for the tensionless
string is not surprising at all. On the other hand, it can be shown,
that in the quantum Virasoro algebra of the usual string the limit
$T\ss 0$ can be taken consistently to obtain the null string gauge algebra
with a vanishing critical dimension \cite{d}.
So what is the correct answer to the
question about the existence of critical dimension for the null string?
In this article we propose a pure technical resolution of the problem.
The right operator ordering is that, which can be applied to higher
dimensions, i.e. to null $p$-branes too. In our case we find two such
orderings and they lead to the absence of critical dimensions for the null
$p$-branes $(p\geq 1)$.

 Here we quantize a model of $p$-branes \cite{1} which initially
do not describe null strings (when $p=1$), because the constraints are second
class. However, it turns out that at the quantum level the constraint
algebra coincides with one of the tensionless string. Checking quantum
consistency of the theory for four different operator orderings
we find $D=26$ for the critical dimension of
the bosonic null string when "string-like" and Weyl orderings are applied.
However, we do not receive any condition on the space-time dimension when
apply "particle-like" and normal ordering. Investigating the case $p>1$,
we observe that the first two orderings are forbidden by the Jacobi
identity. Adopting the last two types of ordering, we reach to the conclusion
that tensionless $p$-branes have no critical dimension for $p>1$.
Because these orderings also apply to the case $p=1$, this conclusion
is valid for all $p=1,2,...$.

 The paper is organized as follows. In section $\bf 2$ we deal with the
 classical theory.
 With the help of the BRST charge, we construct BRST
 invariant hamiltonian and also give the corresponding Lagrangian.
 Then we solve the classical equations of motion and obtain the on-shell
 expressions for the BRST charge $Q$ and for the constraints. Assuming
 periodic boundary conditions, we rewrite all quantities in Fourier modes.
 Section $\bf 3$ is devoted to the quantization of the model. We define the
 renormalized operators and investigate the anomalies in the quantum
 constraint algebra. As a result, we obtain the conditions for quantum
 consistency of the theory for different values of $p$. In section $\bf 4$
 we propose a supersymmetric extension of the model under consideration which
 can be used to describe tensionless super $p$-branes. Finally, in section
 $\bf 5$ we give some comments and conclusions.
\section{\bf Classical theory}
To begin with, we first write down the hamiltonian of the classical model of
$p$-branes with second class constraints \cite{3} proposed in \cite{1}.
It can be cast in the form \cite{2}:
\bq
H_{0}=\int{d^p\sigma\left(\lambda^0\psi_0+\lambda^a\psi_a\right)} ,
\hspace{2cm}
a=1,...,p ,
\eq
where $\lambda^0 , \lambda^a$ are Lagrange multipliers being arbitrary
functions of the time parameter $\tau$ and volume coordinates
$\sigma_1,...,\sigma_p$ . The constraints $\psi_0 , \psi_a$ are
defined by the equalities:
\ba
\psi_0=p^{\alpha}p_{\alpha} + T^2
\hspace{1cm}
,
\hspace{1cm}
\psi_a=\eta_{\alpha\beta}p^\alpha\partial_a x^\beta ,
\\
\nl
\alpha,\beta = 0,1,...,D-2,
\hspace{1cm}
\eta_{\alpha\beta}=diag(-1,1,...,1)
\ea
Here $x^\alpha$ and $p_\alpha$ are canonically conjugated coordinates
and momenta, $\partial_a=\partial/\partial \sigma^a, T=const $.

The algebra of the constraints $(2)$ is given by the Poisson bracket
relations
\ba
\nl
\{\psi_0(\underline{\sigma_1}),
\psi_0(\underline{\sigma_2})\}& = &0 ,
\\
\nl
\{\psi_0(\underline{\sigma_1}),
\psi_a(\underline{\sigma_2})\}& =
&[\psi_0(\ul{\sigma_1})+\psi_0(\ul{\sigma_2})-2T^2]\partial_a \delta^p
(\ul{\sigma_1}-\ul{\sigma_2}) ,
\\
\nl
\{\psi_a(\ul{\sigma_1}),\psi_b(\ul{\sigma_2})\}&=
&[\delta_{a}^{c}\psi_b(\ul{\sigma_1})+
\delta_{b}^{c}\psi_a(\ul{\sigma_2})]
\partial_c\delta^p(\ul{\sigma_1}-\ul{\sigma_2}) ,
\ea
where the notation $\ul{\sigma}=(\sigma_1,...,\sigma_p)$ is used. It
follows from here that the constraints are second class.

Introducing the hamiltonian (1), one has to check the consistency
conditions \cite{3}
\ba
\nl
\{\psi_0,H_0\}\approx 0
\hspace{1cm},
\hspace{1cm}
\{\psi_a,H_0\}\approx 0 ,
\ea
where $\approx$ denotes weak equality, i.e. equality up to constraints.
In the present case these conditions are
\ba
\nl
\p_a\lambda^0 = 0
\h,
\h
\p_a\lambda^a = 0 .
\ea

One of the methods for quantization of dynamical systems with second class
constraints consists in passing to a system with first class constraints only
\cite{5}, and then perform the quantization. To achieve this in our case, we 
enlarge the initial phase space with a new canonical pair $(x_{D-1},p_{D-1})$. 
This allows for transition from initial constraints (2) to the new ones 
\cite{PBMS},\cite{MS}:
\ba
\nl
\vf_0&=&p^{\mu}p_{\mu}=p^{\alpha}p_{\alpha} + p^2_{D-1}
\\
\nl
\vf_a&=&\eta_{\mu\nu}p^{\mu}\p_a x^{\nu}=
\eta_{\alpha\beta}p^{\alpha}\p_a x^{\beta} + p_{D-1}\p_a x_{D-1} .
\ea
$\vf_0$ and $\vf_a$ obey the Poisson bracket algebra
\ba
\nl
\{\varphi_0(\underline{\sigma_1}),
\varphi_0(\underline{\sigma_2})\}& = &0 ,
\\
\{\varphi_0(\underline{\sigma_1}),
\varphi_a(\underline{\sigma_2})\}& =
&[\vf_0(\ul{\sigma_1})+\vf_0(\ul{\sigma_2})]\partial_a \delta^p
(\ul{\sigma_1}-\ul{\sigma_2}) ,
\\
\nl
\{\vf_a(\ul{\sigma_1}),\vf_b(\ul{\sigma_2})\}&=
&[\delta_{a}^{c}\vf_b(\ul{\sigma_1})+
\delta_{b}^{c}\vf_a(\ul{\sigma_2})]
\partial_c\delta^p(\ul{\sigma_1}-\ul{\sigma_2}) ,
\ea
which means, that they are first class quantities. The corresponding
hamiltonian is
\ba
\nl
H=\int{d^p\sigma\left(\mu^0\varphi_0+\mu^a\varphi_a\right)} .
\ea
The Dirac consistency conditions
\ba
\nl
\{\vf_0,H\}\approx{0}
\hspace{1cm},
\hspace{1cm}
\{\vf_a,H\}\approx{0} ,
\ea
do not place any restrictions on the Lagrange multipliers $\mu^0,\mu^a$.

Now, two notes are in order. The first one is that at any moment one can
return to the initial dynamical system by dimensional reduction.
The second is, that the algebra (3) of the constraints $\vf_0,\vf_a$
coincides with the tensionless limit of the usual $p$-brane ones \cite{BN}.
That is why our conclusions about the critical dimensions, arising after
quantization, will be also valid for the tensionless branes.

Following the BFV-BRST method for quantization of constrained systems
\cite{4}, we now introduce for each constraint $\vf_0 , \vf_a$ a pair of
anticommuting ghost variables $(\eta^0,P_0) , (\eta^a,P_a)$
respectively, which are canonically conjugated.Then the BRST charge is
\cite{PLB}
\ba
\nl
Q=\int{d^p\sigma\{\vf_0\eta^0+\vf_a\eta^a+
P_0[(\p_a\eta^a)\eta^0+(\p_a\eta^0)\eta^a]+
P_b(\p_a\eta^b)\eta^a\}}
\ea
and it has the property
\ba
\nl
\{Q,Q\}_{pb}=0
\ea
where $\{.,.\}_{pb}$ is the Poisson bracket in the extended phase space
$(x^\nu,p_\mu;\eta^0,P_0;\eta^a,P_b)$.

In the new phase space, the constraints are given by the following
brackets \cite{6}:
\ba
\nl
\vf_0^{tot}&=&\{Q,P_0\}_{pb}=\vf_0+2P_0\p_a\eta^a+
(\p_a P_0)\eta^a=\vf_0+\vf_0^{gh} ,
\\
\nl
\vf_a^{tot}&=&\{Q,P_a\}_{pb}=\vf_a+2P_0(\p_a\eta^0)+
(\p_a P_0)\eta^0+P_a\p_b\eta^b+
P_b(\p_a\eta^b)+(\p_b P_a)\eta^b=\vf_a+\vf_a^{gh}
\ea
and they are first class. The BRST invariant hamiltonian is \cite{4}
\ba
\nl
H_\chi=\{Q,\chi\}_{pb}
\hspace{1cm},
\hspace{1cm}
\{Q,H_\chi\}_{pb}=0 ,
\ea
where $\chi$ is arbitrary, anticommuting, gauge fixing function. We choose
\ba
\nl
\chi=\Lambda^0\int{d^p\sigma P_0} + \Lambda^a\int{d^p\sigma P_a} ,
\hspace{1cm}
\Lambda^0 , \Lambda^a - const
\ea
and obtain:
\bq
H_\chi=\int{d^p\sigma\bigl[\Lambda^0\vf_0^{tot}+
\Lambda^a\vf_a^{tot}\bigr]} .
\eq

Let us note that additional set of canonically conjugated ghosts
$(\bar{\eta_0},\bar{P^0}) , (\bar{\eta_a},\bar{P^a})$
must be added if we wish to write down the corresponding BRST invariant
Lagrangian. If so, $Q$ and $\chi$ have to be modified in the following
fashion \cite{4,6}
\ba
\nl
\tilde {Q}=Q+\int d^p\sigma(M_0\bar{P^0}+M_a\bar{P^a}) ,
\ea
\ba
\nl
\tilde {\chi}=\chi+\int d^p\sigma[\bar{\eta_0}(\chi^0+
\frac{\rho_1}{2}M^0)+
\bar{\eta_a}(\chi^a+\frac{\rho_2}{2}M^a)] ,
\ea
where $M_0 , M_a$ are the momenta, canonically conjugated to $\mu^0$
and $\mu^a$ respectively, $\chi^0$ and $\chi^a$ are gauge fixing
conditions \cite{7} for $\vf_0$ and $\vf_a$, $\rho_1$ and $\rho_2$ are
parameters. All this results in the Lagrangian density $(\p_\tau=\p/\p\tau)$:
\ba
\nl
L_{\tilde{\chi}}=L+L_{GF}+L_{GH} ,
\ea
where
\ba
\nl
L=(1/4\mu^0)(\p_\tau x-\mu^a\p_a x)^2 ,
\ea
the gauge fixing part is
\ba
\nl
L_{GF}=\frac{1}{2\rho_1}(\p_\tau \mu^0-\chi^0)(\p_\tau \mu_0-\chi_0)+
\frac{1}{2\rho_2}(\p_\tau \mu^a-\chi^a)
(\p_\tau \mu_a-\chi_a)
\ea
and the ghost part is
\ba
\nl
L_{GH}=-\p_\tau\bar{\eta_0}\p_\tau\eta^0-\p_\tau\bar{\eta_a}
\p_\tau\eta^a+\mu^0[2\pu\bar{\eta_0}\pa\eta^a+
(\pa\pu\bar{\eta_0})\eta^a]
\\
\nl
+\mu^a[2\pu\bar{\eta_0}\pa\eta^0+
(\pa\pu\bar{\eta_0})\eta^0+\pu\bar{\eta_b}\pa\eta^b+
\pu\bar{\eta_a}\p_b\eta^b+
(\p_b\pu\bar{\eta_a})\eta^b]
\\
\nl
+\int d^p\sigma'\{\bar{\eta_0}(\sigma')[\{\vf_0,\chi^0(\sigma')\}_{pb}\eta^0
+\{\vf_a,\chi^0(\sigma')\}_{pb}\eta^a]
\\
\nl
+\bar{\eta_a}(\sigma')[\{\vf_0,\chi^a(\sigma')\}_{pb}\eta^0
+\{\vf_b,\chi^a(\sigma')\}_{pb}\eta^b]\} .
\ea

Let us now go back to the hamiltonian picture. The hamiltonian $(4)$ leads
to equations of motion with the following general solution for the bosonic
variables \cite{1}
\ba
\nl
x^\nu&=&y^\nu(\ul{z})+2g(\tau)p^\nu(\ul{z}) ,
\\
\nl
p_\nu&=&p_\nu(\ul{z}) ,
\ea
and for the ghosts \cite{PLB}
\ba
\nl
\eta^0&=&\zeta^0(\ul{z})+g(\tau)\p_a\eta^a(\ul{z}) ,
\\
\nl
P_0&=&P_0(\ul{z}) ,
\\
\nl
\eta^a&=&\eta^a(\ul{z}) ,
\\
P_a&=&\Pi_a(\ul{z})+g(\tau)\p_a P_0(\ul{z}) .
\ea
Here $y^\nu , p_\nu , \zeta^0 , P_0 , \eta^a$ and $\Pi_a$ are
arbitrary functions of the variables $z^a$ ,
\ba
\nl
z^a=\Lambda^a\tau + \sigma^a
\hspace{1cm}
\mbox{and}
\hspace{1cm}
g(\tau)=\Lambda^0\tau .
\ea

On the solutions $(5)$ the BRST charge $Q$ takes the form \cite{PLB}
\ba
\nl
Q^S=\int{d^pz\{\phi_0\zeta^0+\phi_a\eta^a+
P_0[(\p_a\eta^a)\zeta^0+(\p_a\zeta^0)\eta^a]+
\Pi_b(\p_a\eta^b)\eta^a\}} ,
\ea
where $\phi_0=p^2(\ul{z})$ ,
$\phi_a=p_\nu(\ul{z})\p_a y^\nu(\ul{z})$. Now the constraints are
\ba
\nl
\phi_0^{tot}(\ul{z})=\{Q^S,P_0(\ul{z})\}_{pb}
\h,
\h
\phi_a^{tot}(\ul{z})=\{Q^S,\Pi_a(\ul{z})\}_{pb} ,
\ea
and they are connected with $\vf_0^{tot} , \vf_a^{tot}$
by the equalities
\ba
\nl
\vf_0^{tot}(\ul{z})=\phi_0^{tot}(\ul{z})
\h,
\h
\vf_a^{tot}=\phi_a^{tot}(\ul{z})+
g(\tau)\p_a\phi_0^{tot}(\ul{z}) .
\ea

From now on, we confine ourselves to the case of periodic boundary conditions
when our phase-space variables admit Fourier series expansions.
Let us denote the Fourier components of $y^\nu, p^\nu, \zeta^0, P_0,
\eta^a$ and $\Pi_a$ with $x^{\nu}_{\ul k}, p^{\nu}_{\ul k},
c_{\ul k}, b_{\ul k}, \bar{c}^{a}_{\ul k}$ and $\bar{b}_{a ,\ul k}$
respectively. For the zero modes of $p^\nu$ and $x^\nu$, we introduce
the notations
\ba
\nl
P^\mu=(2\pi)^pp^{\mu}_{\ul 0} \h ,
\h q^\nu=\frac{-i}{(2\pi)^p}x^{\nu}_{\ul 0} .
\ea
Then we have the following non-zero Poisson brackets:
\ba
\nl
\{P^\mu,q^\nu\}_{pb}&=&-\eta^{\mu\nu},
\\
\nl
\{p^{\mu}_{\ul k},x^{\nu}_{\ul n}\}_{pb}&=&
-i\eta^{\mu\nu}\delta_{\ul k +\ul n,\ul 0},
\\
\{c_{\ul k},b_{\ul n}\}_{pb}&=&-i\delta_{\ul k +\ul n,\ul 0},
\\
\nl
\{\bar{c}^{a}_{\ul k},\bar{b}_{b,\ul n}\}_{pb}&=&
-i\delta^{a}_{b}\delta_{\ul k +\ul n,\ul 0} .
\ea

The Fourier expansions for the constraints $\phi_0^{tot}$ and
$\phi_a^{tot}$ are
\ba
\nl
\phi_0^{tot}(\ul{z})=\frac{1}{(2\pi)^p}\sum_{\ul{m}\in Z^p}
C_{\ul{m}}^{tot}e^{-i\ul{m}\ul{z}}
\h,
\h
\phi_a^{tot}(\ul{z})=\frac{1}{(2\pi)^p}\sum_{\ul{m}\in Z^p}
D_{a,\ul m}^{tot}e^{-i\ul{m}\ul{z}} .
\ea
Here
\bq
C_{\ul{n}}^{tot}=i\{Q^S,b_{\ul{n}}\}_{pb}=
C_{\ul{n}}+C_{\ul{n}}^{gh}
\h,
\h
D_{a,\ul n}^{tot}=i\{Q^S,\bar{b}_{a,\ul n}\}_{pb}=
D_{a,\ul n}+D_{a,\ul n}^{gh}
\eq
where
\ba
\nl
Q^S&=&\sum_{\ul{n}\in Z^p}\{[C_{\ul{n}}+(1/2)C_{\ul{n}}^{gh}]c_{-\ul{n}}+
[D_{a,\ul n}+(1/2)D_{a,\ul n}^{gh}]\bar{c}_{-\ul{n}}^a\} ,
\\
\nl
C_{\ul n}&=&(2\pi)^p\sum_{\ul k\in Z^p}p^{\nu}_{\ul k}p_{\nu,\ul n-\ul k},
\\
D_{a,\ul n}&=&-\sum_{\ul k\in Z^p}(n_a-k_a)p^{\nu}_{\ul k}
x_{\nu,\ul n-\ul k} ,
\\
\nl
C^{gh}_{\ul n}&=&\sum_{\ul k\in Z^p}(n_a - k_a)b_{\ul n +\ul k}
\bar{c}^{a}_{-\ul k} ,
\\
\nl
D^{gh}_{a,\ul n}&=&\sum_{\ul k\in Z^P}[(n_a - k_a)
b_{\ul n +\ul k}c_{-\ul k} + (\delta_{a}^{c}n_b -
\delta_{b}^{c}k_a)\bar{b}_{c,\ul n +\ul k}
\bar{c}^{b}_{-\ul k}]
\ea

Using expressions $(6)$ to $(8)$, one obtains that the algebra of the
total generators $(7)$ is given by
\ba
\nl
\{C_{\ul{n}}^{tot},C_{\ul{m}}^{tot}\}_{pb}&=&0 ,
\\
\nl
\{C_{\ul{n}}^{tot},D_{a,\ul m}^{tot}\}_{pb}&=
&-i(n_{a}-m_{a})C_{\ul{n}+\ul{m}}^{tot} ,
\\
\{D_{a,\ul n}^{tot} , D_{b,\ul m}^{tot}\}_{pb}&=
&-i(\delta^{c}_{a}n_{b}-
\delta^{c}_{b}m_{a})D_{c,\ul n +\ul m}^{tot} .
\ea

\section{\bf Quantization}

Going to the quantum theory according to the rule $i\{.,.\}_{pb}\rightarrow$
(anti)commutator, we define $Q^S$ by introducing the renormalized operators
$(\alpha,\beta_a - const)$
\bq
C_{\ul{n}}^{tot}=C_{\ul{n}}+C_{\ul{n}}^{gh}-\alpha\delta_{\ul{n},\ul{0}}
\h ,
\h
D_{a,\ul{n}}^{tot}=D_{a,\ul{n}}+D_{a,\ul{n}}^{gh}-
\beta_a\delta_{\ul{n},\ul{0}}
\eq
and postulating \cite{PLB}
\ba
\nl
Q^S=\sum_{\ul{n}\in Z^p}:\{[C_{\ul{n}}+(1/2)C_{\ul{n}}^{gh}
-\alpha\delta_{\ul{n},\ul{0}}]c_{-\ul{n}}
\\
\nl
+[D_{a,\ul{n}}+(1/2)D_{a,\ul{n}}^{gh}-
\beta_a\delta_{\ul{n},\ul{0}}]\bar{c}_{-\ul{n}}^a\}: ,
\ea
where :...: stands for operator ordering and in
$C_{\ul{n}},...,D_{a,\ul{n}}^{gh}$ operator ordering is also assumed.

Let us turn to the question about the critical dimensions which might
appear in the model under consideration. As is well known, the critical
dimension arises as a necessary condition for nilpotency of the BRST
charge operator. In turn, this is connected with the vanishing of the
central charges in the quantum constraint algebra. Because of that,
we are going to find out the central terms which appear in our
quantum gauge algebra for different values of $p$ (the most general form
of central extension, which is compatible with the Jacobi identities is
written in the Appendix).

We start with the case $p=1$, which corresponds to a closed string.
In this case $a=b=1$ and one defines the operator ordering with
respect to $p_{-n}^\nu,...,\bar{c}_{-n}$ and $p_n^\nu,...,\bar{c}_n , (n>0)$,
so that
\ba
\nl
p_{-n}^{\nu}\mid 0> = ... =\bar{c}_{-n}\mid 0> = 0
\h ,
\h
<0\mid p_{n}^{\nu} = ... =<0\mid \bar{c}_{n} = 0.
\ea
We call this ordering "$string-like$". Using the explicit expressions for the
constraints $(8)$, one obtains that central terms appear in the commutators
$[D_n,D_m] , [D_n^{gh},D_m^{gh}]$ and they are respectively
\ba
\nl
c=(D/6)(n^2-1)n\delta_{n+m,0}
\h ,
\h
c^{gh}=-(1/3)(13n^2-1)n\delta_{n+m,0} .
\ea
Therefore, the quantum constraint algebra has the form
\ba
\nl
[C_n^{tot},C_m^{tot}]&=&0 ,
\\
\nl
[C_n^{tot},D_m^{tot}]&=&(n-m)C_{n+m}^{tot}+2\alpha n\delta_{n+m,0} ,
\\
\nl
[D_n^{tot},D_m^{tot}]&=&(n-m)D_{n+m}^{tot}+
(1/6)[(D-26)n^2+(12\beta-D+2)]n\delta_{n+m,0} .
\ea
This means that the conditions for the nilpotency of the BRST charge
operator $Q^S$ are
\ba
\nl
(D-26)n^2+(12\beta-D+2)=0
\h ,
\h
\alpha=0 ,
\ea
which leads to the well known result $D=26,\beta=2$. Obviously,
this reproduces one of the basic features of the quantized tensionful closed
bosonic string - its critical dimension.

Going to the case $p>1$, one natural generalization of the creation and
annihilation operators definition is
\ba
\nl
p_{\ul{n}}^\nu\mid 0>_a=0 ,
\h
_a<0\mid p_{-\ul{n}}^\nu=0 ,
\h
\mbox{for}
\h
\sum_{a=1}^{p}n_{a} > 0
\ea
and analogously for the operators $x_{\ul{n}}^\nu,...,\bar{c}_{\ul{n}}^a$.
However, it turns out that such definition does not agree with the Jacobi
identities for the quantum constraint algebra (except for $p=1$).
That is why, we introduce the creation $(+)$
and annihilation $(-)$ operators in the following way \cite{PLB}
\bq
p_{\ul{n}}^\nu=(1/\sqrt{2})(p_{\ul{n}}^{\nu+}+p_{-\ul{n}}^{\nu-}) ,
...,\bar{c}_{\ul{n}}^a=(1/\sqrt{2})(\bar{c}_{\ul{n}}^{a+}+
\bar{c}_{-\ul{n}}^{a-})
\eq
and respectively new vacuum states
\ba
\nl
p_{\ul n}^{\nu-}\mid vac> = ... = \bar{c}_{\ul n}^{a-}\mid vac> = 0
\h ,
\h
<vac\mid p_{\ul n}^{\nu+} = ... = <vac\mid \bar{c}_{\ul n}^{a+} = 0 .
\ea
This choice of the creation and annihilation operators corresponds to the
representation of all phase-space variables $p^\nu,...,\bar{c}^a$ as
sums of frequency parts which are conjugated to each other and satisfy
the same equation of motion as the corresponding dynamical variable.

By direct computation one shows, that with operator product defined with
respect to the introduced creation and annihilation operators $(11)$
(we shall refer to as "$normal$ $ordering$"), the central extension of the
algebra of the gauge generators $(10)$ does not appear, i.e. $\alpha=0 ,
\beta_a=0$. Consequently, the BRST charge operator $Q^S$ is automatically
nilpotent in this case and there is no restriction on the dimension of the
background space-time for $p>1$.

The impossibility to introduce a $string-like$ operator ordering when $p>1$
leads to the problem of finding those operator orderings which are possible
for $p=1$ as well as for $p>1$. First of all, we check the consistency of the
(already used for $p>1$)
$normal$ $ordering$ for $p=1$. It turns out that it is consistent, but now
critical dimension for the null string does not appear. The same result -
absence of critical dimension for every value of $p$, one obtains when uses
the so called $particle-like$ operator ordering. Now the $ket$ vacuum is
annihilated by momentum-type operators and the $bra$ vacuum is annihilated by
coordinate-type ones:
\ba
\nl
p_{\ul n}^{\mu}\mid 0>_M &=& b_{\ul n}\mid 0>_M = \bar{b}_{\ul n}\mid 0>_M = 0 ,
\\
\nl
_C<0\mid x_{\ul n}^{\mu} &=& _C<0\mid c_{\ul n} = _C<0\mid \bar{c}_{\ul n} = 0
\hspace{.5cm} ,
\hspace{.5cm}
\forall \ul n \in Z^p .
\ea
Further, we check the case when $Weyl$ $ordering$ is applied. Now it turns out,
that in the null string case ($p=1$) this leads to critical dimension $D=26$,
but for the null brane ($p>1$) this ordering is inconsistent, as was the
$string-like$ one.

As a final result, we have four type of operator orderings checked. Two of
them are valid for the string as well as for the brane and then we do not
receive any critical dimension. The other two type of ordering give
critical dimension $D=26$ for the string and are not applicable for the brane.
Our opinion is that the right operator ordering is the one applicable for
all $p=1,2,...$. In other words, our viewpoint is that neither null
strings nor null branes have critical dimensions.
The same point of view is presented in \cite{e}.

Let us spend some more words about the impossibility to introduce at $p>1$ an
operator ordering which at $p=1$ gives critical dimension. This is connected
with the fact that the constraint algebra, as is shown in the Appendix,
does not possess non-trivial central extension when $p>1$
(see also \cite{2}, \cite{e}). As a matter of fact, the string
critical dimension appears in front of $n^3$, i.e. in the
non-trivial part of the constraint algebra central extension, which can not
be taken away by simply redefining the generators $D_n$, in contrast to the
trivial part $\sim n$. Because of the nonexistence of non-trivial central
extension when $p>1$, any critical dimension arising is impossible in view
of the Jacobi identities. Therefore, if the quantum null brane
constraint algebra is given by (up to trivial central extensions)
\ba
\nl
[C_{\ul{n}}^{tot},C_{\ul{m}}^{tot}]&=&0 ,
\\
\nl
[C_{\ul{n}}^{tot},D_{a,\ul m}^{tot}]&=&
(n_{a}-m_{a})C_{\ul{n}+\ul{m}}^{tot} ,
\\
\nl
[D_{a,\ul n}^{tot} , D_{b,\ul m}^{tot}]&=&
(\delta^{c}_{a}n_{b}-
\delta^{c}_{b}m_{a})D_{c,\ul n +\ul m}^{tot} ,
\ea
then the latter has no critical dimension and exists in
any D-dimensional space-time, when embedding of the $p+1$-
dimensional worldvolume of the $p$-brane is possible there.

Finally, we pay attention to the fact that in every one of the $p$ subalgebras
(at fixed $a$) of the constraint algebra, one can obtain non-trivial central
extension and consequently - critical dimension (see Appendix). For example,
taking $string-like$ or $Weyl$ $ordering$, one derives $D=25+p$, which
appears to be critical dimension for the tensile $p$-brane \cite{Marq},
\cite{e}. However, the considered quantum dynamical system is described by
the $\ul{full}$ constraint algebra, where only trivial central extensions
are possible.
\section{\bf Supersymmetrization}
It turns out that the model described in the previous sections can be
generalized to include also spinorial degrees of freedom. This generalization
is not straightforward, but the resulting dynamical system may be viewed as
generated by its bosonic part, which in terms of constraints is equivalent
to a system with Poisson bracket relations, given by $(9)$, i.e. equivalent
to the null bosonic brane. This new model possesses space-time supersymmetry
and is characterized by the following classical first class constraints
\ba
\nl
\{T_0(\ul \sigma_1),T_0(\ul \sigma_2)\}&=&0,
\\
\nl
\{T_0(\ul \sigma_1),T_{\alpha}^{A}(\ul \sigma_2)\}&=&0 ,
\\
\nl
\{T_0(\ul \sigma_1),T_{a}^{A}(\ul \sigma_2)\}&=&
[T_0(\ul \sigma_1) + T_0(\ul \sigma_2)]
\pa \delta^{p}(\ul \sigma_1 - \ul \sigma_2) ,
\\
\{T_{a}^{A}(\ul \sigma_1),T_{b}^{B}(\ul \sigma_2)\}&=&
\delta^{AB}[\delta_{a}^{c}T_{b}^{B}(\ul \sigma_1) +
\delta_{b}^{c}T_{a}^{A}(\ul \sigma_2)]\p_c
\delta^{p}(\ul \sigma_1-\ul \sigma_2) ,
\\
\nl
\{T_{a}^{A}(\ul \sigma_1),T_{\alpha}^{B}(\ul \sigma_2)\}&=&
\delta^{AB}[T_{\alpha}^{A}(\ul \sigma_1)+T_{\alpha}^{A}(\ul \sigma_2)]
\pa \delta^{p}(\ul \sigma_1-\ul \sigma_2) ,
\\
\nl
\{T_{\alpha}^{A}(\ul \sigma_1),T_{\beta}^{B}(\ul \sigma_2)\}&=&
-2i\delta^{AB}\hat{P}_{\alpha\beta}T_0(\ul \sigma_1)
\delta^{p}(\ul \sigma_1-\ul \sigma_2) ,
\\
\nl
\hat{P}_{\alpha\beta}&=&P_{\mu}\sigma_{\alpha\beta}^{\mu} .
\ea

Comparing the above equalities with the spinning string and superstring
constraint algebras, we
conclude that they can be regarded as possible tensionless limit of the
super $p$-brane case. However, this supersymmetric model will be considered
in detail in a separate paper. Here we only note, that Poisson brackets in
$(12)$ give the naive version of the constraint algebra. Actually, there is
a set of generators with which $(12)$ must be enlarged.

\section{\bf Comments and Conclusions}

 In this paper we present the results on the quantization of the restricted
$p$-brane \cite{1} reported in \cite{PLB}. On the other hand, we
investigate the connection between the appearance of critical dimensions and
different operator orderings for $p=1,2,...$ .

 The quantization of the restricted $p$-brane in \cite{PLB} is alternative to
the one given in \cite{MS}. The latter is based on a previous work
\cite{PBMS} on the quantization of the restricted string and treat
asymmetrically the constraints $p^{\nu}\p_a x_{\nu}=0$ for $a=1$ and
$a=2,3,...,p$. In \cite{PLB} and here, we consider all these constraints
on equal footing.

 The observation, that there is an operator ordering which is valid
$\forall p\in Z_{+}$ and another one, which is admissable only for $p=1$
\cite{PLB}, leads to the problem of finding those orderings which are
possible for every positive integer value of $p$. We applied here four
types of operator orderings and we establish that two of them
($normal$ $ordering$ and $particle-like$ $ordering$)
are admissable $\forall p\in Z_{+}$, but the other two
($string-like$ and $Weyl$ $ordering$) are admissable only for $p=1$.
The fact, that the latter two orderings lead to appearance of critical
dimension, and the former two do not, is a consequence of the constraint
algebra property to have non-trivial central extension only for $p=1$.
On the other hand, the obtained nontrivial central extensions
of the Virasoro type for some of it subalgebras, provide an explanation why
the critical dimensions $D=25+p, p=1,2,...$ \cite{Marq},\cite{e}, re-derived
also here, can emerge. However, our claim is, that the critical dimensions
appearing in the subalgebras, must not be considered as such for the
given model as a whole. The model is represented by the full constraint
algebra, which does not possess non-trivial central extension for $p\geq 2$.

 Since after BFV-BRST quantization our constraint algebra coincides with
the null tension limit of the usual $p$-brane algebra \cite{BN}, we deduce
that the upper conclusions are valid for the tensionless $p$-branes also.
This lead us to the proposition of the rule: the right operator orderings
in the case of null string ($p=1$) are those, which are admissable in the
$p>1$ case too.

\vspace*{.5cm}
{\bf Acknowledgments}

The author would like to thank A. Pashnev, M. Stoilov and D. Stoyanov
for the useful discussions, and B. Dimitrov for critical reading of the
manuscript. It is also a pleasure to thank F. Lizzi, A. Nicolaidis,
P. Porfyriadis and P. Saltsidis for the given information about their
related papers.

\vspace*{1cm}
{\Large{\bf Appendix}}
\vspace*{.5cm}

Here we briefly comment on the possible central extensions of the
algebra, given by the commutators:
\ba
\nl
[A_{\ul n},A_{\ul m}]&=&0 ,
\\
\nl
[A_{\ul n},B_{a,\ul m}]&=&(n_a - m_a)A_{\ul n +\ul m} ,
\\
\nl
[B_{a,\ul n},B_{b,\ul m}]&=&(\delta_{a}^{c} n_b - \delta_{b}^{c} m_a)
B_{c,\ul n +\ul m}
\h ,
\h
(a,b=1,2,...,p) .
\ea
To begin with, we modify the right hand sides of the upper equalities in
the following way:
\ba
\nl
[A_{\ul n},A_{\ul m}]&=&d(\ul n,\ul m)
\\
\nl
[A_{\ul n},B_{a,\ul m}]&=&(n_a - m_a)A_{\ul n +\ul m}+d_a (\ul n,\ul m) ,
\\
\nl
[B_{a,\ul n},B_{b,\ul m}]&=&(\delta_{a}^{c} n_b - \delta_{b}^{c} m_a)
B_{c,\ul n +\ul m}+d_{ab}(\ul n,\ul m) .
\ea
Checking the Jacobi identities, involving the triplets $(A,A,B),(A,B,B)$ and
$(B,B,B)$,one shows that there are only trivial solutions for
$d(\ul n,\ul m)$, $d_{a}(\ul n,\ul m)$ and $d_{ab}(\ul n,\ul m)$. Namely,
\ba
\nl
d(\ul n,\ul m)&=&0
\h ,
\h
d_{a}(\ul n,\ul m)=(n_a - m_a)f(\ul n +\ul m) ,
\\
\nl
d_{ab}(\ul n,\ul m)&=&(\delta_{a}^{c}n_b - \delta_{b}^{c}m_a)
g_{c}(\ul n + \ul m) ,
\ea
where $f, g_a$ are arbitrary functions of their arguments. In particular,
there exist the solutions
\ba
\nl
d_{a}(\ul n,\ul m)&=&2\alpha n_a \delta_{\ul n +\ul m,\ul 0}
\h ,
\h
\alpha=const ,
\\
\nl
d_{ab}(\ul n,\ul m)&=&(\beta_{a}n_b + \beta_{b}n_a)\delta_{\ul n +\ul m,\ul 0}
\h ,
\h
\beta_a = const ,
\ea
which might appear because of the operator ordering in $A_{\ul n}$ and
$B_{a,\ul n}$. However, there are $p$ subalgebras with non-trivial central
extensions (no summation over $a$):
\ba
\nl
[A_{\ul n},A_{\ul m}]&=&0 ,
\\
\nl
[A_{\ul n},B_{a,\ul m}]&=&(n_a - m_a)A_{\ul n +\ul m}+
(q_{a}n_{a}^2+r_{a})n_{a}\delta_{\ul n +\ul m,\ul 0} ,
\\
\nl
[B_{a,\ul n},B_{a,\ul m}]&=&(n_a - m_a)
B_{a,\ul n +\ul m}+(s_{a}n_{a}^2 + t_{a})n_{a}\delta_{\ul n +\ul m,\ul 0} ,
\hspace{1cm}
q_{a}, r_{a}, s_{a}, t_{a} - const .
\ea
When $p=1$, there is one such subalgebra and it coincides with the full
algebra.


\begin{thebibliography}{17}

\bibitem{A} H. de Vega, A. Nicolaidis, Phys. Lett. {\bf B295} (1992) 214;

O. Ganor, A. Hanani, Nucl. Phys. {\bf B474} (1996) 122, hep-th/9602120;

C. Lousto, N. Sanches, Phys. Rev. {\bf D54} (1996) 6399, gr-qc/9605015;

O. Ganor, Nucl. Phys. {\bf B489} (1997) 95, hep-th/9605201;

A. Hanani, I. Klebanov, Nucl. Phys. {\bf B482} (1996) 105, hep-th/9606136;

O. Ganor, hep-th/9607092;

K. Davis, hep-th/9609005;

P. Mayr, Nucl. Phys. {\bf B494} (1997) 489, hep-th/9610162;

M. Dabrowski, A. Larsen, Phys. Rev. {\bf D55} (1997) 6409, hep-th/9610243;

B. Jensen, U. Lindstr$\ddot{o}$m, Phys. Lett. {\bf B398} (1997) 83,
hep-th/9612213;

P. Porfyriadis, D. Papadopoulos, hep-th/9707183;

K. Ilienko, A. Zheltukhin, hep-th/9708024.
\bibitem{B} F. Lizzi, B. Rai, G. Sparano, A. Srivastava, Phys. Lett.
{\bf B182} (1986) 326;

A. Karlhede, U. Lindstr${\ddot{o}}$m, Class. Quant. Grav. {\bf 3} (1986) L73;

R. Amorim, J. Barcelos-Neto, Z. Phys. {\bf C38} (1988) 643;

J. Barcelos-Neto, M. Ruiz-Altaba, Phys. Lett. {\bf B228} (1989) 193;

I. Bandos, A. Zheltukhin, Sov. J. Nucl. Phys. {\bf 50} (1989) 556;

H. de Vega, I. Giannakis, A. Nicolaidis, Mod. Phys. Lett. {\bf A10} (1995)
2479, hep-th/9412081.
\bibitem{C} J. Gamboa, C. Ramirez, M. Ruiz-Altaba, Phys. Lett.
{\bf B225} (1989) 335;

R. Amorim, J. Barcelos-Neto, Phys. Lett. {\bf 253} (1991) 313.
\bibitem{d} F. Lizzi, Mod. Phys. Lett. {\bf A9} (1994) 1495, hep-th/9404148;

A. Nicolaidis, J. Paschalis, P. Porfyriadis, hep-th/9702185.
\bibitem{1} D. Stoyanov, Mod. Phys. Lett. {\bf A4} (1989) 1287.
\bibitem{3} P. A. M. Dirac, Lectures on Quantum Mechanics
(Yeshiva University, New York, 1964).
\bibitem{2} M. Stoilov, D. Stoyanov, J. Phys. {\bf A23} (1990) 5925.
\bibitem{5} L. Faddeev, S. Shatashvili, Phys. Lett. {\bf B167} (1986) 225;

E. Egoryan, R. Manvelyan, Theor. Math. Phys. {\bf 94} (1993) 241.

\bibitem{PBMS} P. Bozhilov, M. Stoilov, Phys. Lett. {\bf B293} (1992) 335.
\bibitem{MS} M. Stoilov, Phys. Lett. {\bf B316} (1993) 80;
\bibitem{BN} B. Barbashov and V. Nesterenko, Introduction to the relativistic
string theory, (World Scientific, Singapore, 1990);

\bibitem{4} E. Fradkin, G. Vilkovisky, Phys. Lett. {\bf B55} (1975) 224;

I. Batalin, G. Vilkovisky, Phys. Lett. {\bf B69} (1977) 309;

E. Fradkin, T. Fradkina, Phys. Lett. {\bf B72} (1978) 343;

M. Henneaux, Phys. Rep. {\bf 126} (1985) 1.

\bibitem{PLB} P. Bozhilov, BRST quantization of a bosonic $p$-brane model, in:
Proc. of $25^{th}$ Anniversary Conference of the "Konstantin Preslavsky" Univ.
of Shoumen, 30.10-02.11.1996, Shoumen, Bulgaria.
\bibitem{6} M. Henneaux, Phys. Lett. {\bf B120} (1983) 179;

K. Fujikawa, J. Kubo, Phys. Lett. {\bf B199} (1987) 75.
\bibitem{7} R. Marnelius, Phys. Lett. {\bf B99} (1981) 467;

R. Marnelius, Acta Phys. Polon. {\bf B13} (1982) 669.
\bibitem{e} P. Saltsidis, Phys. Lett. {\bf B401} (1997) 21, hep-th/9702081;
\bibitem{Marq} U. Marquard, M. Scholl, Phys. Lett. {\bf B209} (1988) 434;
{\bf B227} (1989) 227;

\end{thebibliography}
\end{document}